\documentclass[12pt]{article}

\usepackage{graphicx}
\usepackage{times}
\usepackage{color}      % use if color is used in text

\newenvironment{sciabstract}{%
\begin{quote} \bf}
{\end{quote}}

\title{Tuning from failed superconductor to failed insulator with magnetic field}

\author
{Yangmu Li,$^{1}$ J. Terzic,$^{2}$ P. G. Baity,$^{2}$ Dragana Popovi\'{c},$^2$ G. D. Gu,$^1$ Qiang Li,$^1$\\
 A. M. Tsvelik,$^1$ J. M. Tranquada$^{1\ast}$\\
\\
\normalsize{$^{1}$Condensed Matter Physics and Materials Science Division, }\\
\normalsize{Brookhaven National Laboratory, Upton, New York 11973, USA}\\
\normalsize{$^{2}$National High Magnetic Field Laboratory,}\\
\normalsize{Florida State University, Tallahassee, Florida 32310, USA}\\
\\
\normalsize{$^\ast$To whom correspondence should be addressed; E-mail:  jtran@bnl.gov.}
}

\date{}

\begin{document} 


\baselineskip24pt


\maketitle


\begin{sciabstract}
  Do charge modulations compete with electron pairing in high-temperature copper-oxide superconductors?  We investigated this question by suppressing superconductivity in a stripe-ordered cuprate compound at low temperature with high magnetic fields.  With increasing field, loss of three-dimensional superconducting order is followed by reentrant two-dimensional superconductivity and then an ultra-quantum metal phase.  Circumstantial evidence suggests that the latter state is bosonic and associated with the charge stripes.  These results provide experimental support to the theoretical perspective that local segregation of doped holes and antiferromagnetic spin correlations underlies the electron-pairing mechanism in cuprates. 
\end{sciabstract}

% In setting up this template for *Science* papers, we've used both
% the \section* command and the \paragraph* command for topical
% divisions.  Which you use will of course depend on the type of paper
% you're writing.  Review Articles tend to have displayed headings, for
% which \section* is more appropriate; Research Articles, when they have
% formal topical divisions at all, tend to signal them with bold text
% that runs into the paragraph, for which \paragraph* is the right
% choice.  Either way, use the asterisk (*) modifier, as shown, to

\newpage

\paragraph¥{ INTRODUCTION}\ \ 

A variety of electronic orders have been proposed and/or observed to exist within the generic phase diagram of copper-oxide high-temperature superconductors \cite{keim15}, and most are commonly viewed as competing with the superconducting order.  In the case of charge-stripe order \cite{oren00}, however, it has been argued that electron pairing and superconducting order can actually be intertwined with the charge modulations \cite{frad15}.   Moreover, increasingly powerful numerical calculations indicate that charge stripes are a natural consequence of doping holes into a correlated antiferromagnetic insulator \cite{zhen17,huan17} and can exhibit superconducting correlations \cite{whit09,corb14}.  While charge stripe order is readily observed within the CuO$_2$ planes of compounds such as La$_{2-x}$Ba$_x$CuO$_4$ (LBCO) \cite{huck11}, experimentally establishing a positive connection with superconductivity has been challenging.   One theoretical expectation for a superconductor based on stripes is that the magnitude of the energy gap associated with pair correlations within charge stripes should be much greater than the energy associated with coherent coupling between the stripes \cite{carl00}, and here we test the implications of this picture.  We use a transverse magnetic field first to decouple the superconducting planes in LBCO with $x=0.125$ and then to destroy the superconducting order within the planes.  We discover an ultra-quantum metal phase that we argue is not easily explained by conventional fermionic quasiparticles, in sharp contrast to the high-field behavior in YBa$_2$Cu$_3$O$_{6+x}$ at a similar hole concentration \cite{seba15}.  We conclude that it provides circumstantial evidence that charge stripes can exhibit robust pairing correlations.

What is already known about LBCO? The spin and charge stripe orders have been studied previously by neutron and x-ray diffraction techniques in magnetic fields perpendicular to the CuO$_2$ planes up to approximately 10 T.  For hole concentrations close to $x=0.125$, the field enhances the stripe order parameters \cite{huck13}, while right at $x=0.125$ it simply increases the charge-stripe correlation length \cite{kim08b}.  [Field-induced charge order has been reported in YBa$_2$Cu$_3$O$_{6+x}$ in fields up to 28~T \cite{gerb15}.]  While the bulk $T_c$ is suppressed to  $\sim 5$~K in zero field for $x=0.125$, a substantial decrease in the $ab$-plane resistivity already occurs below $\sim40$~K, with additional evidence for a transition to two-dimensional (2D) superconducting order near 16~K \cite{li07}.  One would expect Josephson coupling between neighboring layers to induce 3D superconducting order as soon as 2D superconductivity develops; the apparent frustration of the interlayer Josephson coupling  has been explained in terms of a proposed pair-density-wave (PDW) state, which involves a strong pairing amplitude on the 1D charge stripes but with opposite signs of the pair wave function on neighboring stripes \cite{hime02,berg07}. In LBCO with $x=0.095$, where the 3D superconductivity is more robust, a magnetic field perpendicular to the CuO$_2$ planes causes a decoupling into 2D superconducting layers \cite{steg13}. 

\paragraph¥{RESULTS}\ \ 

Following the first successful growth of a La$_{2-x}$Ba$_x$CuO$_4$ single crystal with $x\sim1/8$ and observation of stripe order by neutron diffraction \cite{fuji04}, considerable time was devoted to growing a series of large crystals with $0.095\le x\le 0.155$ in steps of $\Delta x=0.02$ utilizing infrared image furnaces and the traveling solvent floating-zone technique \cite{tana98}.  X-ray and neutron diffraction studies have demonstrated that the structural transitions and lattice parameters are quite sensitive to $x$, providing valuable and reliable measures of stoichiometry \cite{huck11}.  The crystals of $x=0.125$ were originally grown for a neutron-scattering study of the spin dynamics \cite{tran04}, and have since been characterized by transport \cite{li07,tran08} and optical conductivity \cite{home12}.  The crystals used in the present study, described further in the Materials and Methods section, were from the same growth.

A partial summary of our observations is presented in Fig.~1A as a color contour map of electrical resistivity as a function of temperature and magnetic field. The electrical resistivity is presented throughout this work in terms of the resistance per CuO$_2$ sheet (sheet resistance), $R_s = \rho_{ab}/d$, where $d=6.6$~\AA\ is the layer separation, with the magnitude in units of the quantum of resistance for electron pairs, $R_Q = h/(4e^2) = 6.45$~k$\Omega$, where $h$ is Planck's constant and $e$ is the electron charge.  An increase of $R_s$ through $R_Q$ is associated with the localization of electron pairs, as observed in the carrier-density-tuned superconductor-insulator transition for La$_{2-x}$Sr$_x$CuO$_4$ thin films \cite{boll11}. At low temperatures with increasing magnetic field $H$, we observe a progression from 3D superconducting order, through a reentrant 2D superconducting order, to an anomalous high-field metallic state with a sheet resistance saturated at $R_s \approx 2R_Q$.  We denote this unanticipated state as an ultra-quantum metal (UQM): it is a metal because the resistance appears not to change as the temperature is reduced towards zero, and it is ``ultra-quantum'' because the magnitude of $R_s$ cannot be explained by the usual semiclassical models.  Another significant observation is that the Hall coefficient $R_{\rm H}$ is negligible below 15~K for the full field range [see Fig.~1B and Fig.~S2 in the Supplementary Materials (SM)], a behavior expected in a superfluid of bosonic Cooper pairs, but that survives even in the UQM phase.

Before proceeding to the high-field results, we first consider the temperature dependence of $R_s$ in zero-field.  It is known from previous work that the anisotropy between the $c$-axis resistivity, $\rho_c$, and the in-plane resistivity, $\rho_{ab}$, is $\sim10^4$ near 40~K, and rises to $\sim10^5$ below the onset of 2D superconducting correlations \cite{li07}.  This large anisotropy makes the measurement of $\rho_{ab}$ quite sensitive to sample imperfections.  The inset of Fig.~2 shows the sample configuration; the current is directly injected into the CuO$_2$ planes and the voltage probes contact the edges of those same planes.  If there is a slight misorientation of the crystal such that the $c$-axis is not precisely perpendicular to the current direction, then a small fraction of the current path will be along the $c$ axis, making the measurement sensitive to $\rho_c$.  With that preface, consider the measurements of $R_s$ shown in Fig.~2 for two samples, A and B, with results from voltage contacts on both sides of A, labelled A1 and A2.   The responses observed for A1 and A2 are consistent with previous work \cite{li07}: a slight jump in $R_s$ and change in slope at 56~K, corresponding to a well-known structural transition and the coincident charge-stripe ordering \cite{fuji04,huck11}, and a large decrease below $\sim36$~K indicating the mean-field transition to 2D superconductivity.  In contrast, sample B shows a distinct behavior, with a larger magnitude of $R_s$ at high temperature, a significant enhancement of resistance on cooling, and a peak in resistivity at 29~K. Such behavior, which resembles that of $\rho_c$ \cite{li07} and has been reproduced in other samples, indicates a contribution from $\rho_c$ consistent with a misorientation of $\sim1.5^\circ$.  (See the SM for further details.)  The differences between samples A and B, while unintended, provide valuable information regarding the field-dependent behavior.

Next we consider the magnetic-field dependence of  $R_s$ measured at various fixed temperatures.  Data for A1, A2, and B up to 35~T are shown in Fig.~3A-C.  At low temperature, we find that the data collapse with a simple {\it ad hoc} scaling determined by characteristic fields $H_{\rm 3D}$ and $H_{\rm 2D}$.   We define $H_{\rm 3D}$ to be the field at which we detect an initial onset of finite resistivity, indicating loss of 3D superconducting order.  $H_{\rm 2D}$ corresponds to the mid-range of the reentrant 2D superconductivity, as discussed below.  (In practice, it was determined by the corresponding local maximum in $R_s$ for A2.)  Figure~3D-F show $R_s$ plotted vs.\ $(H-H_{\rm 2D})/\Delta H$, where $\Delta H = H_{\rm 2D}-H_{\rm 3D}$.   Here we see that all three data sets are identical up to $H_{\rm 3D}$.  For A1, $R_s$ then returns to zero for a finite range of field.  In the same region, B shows a rise to above $4R_Q$, consistent with insulating behavior in the $\rho_c$ contribution due to putative PDW order and 2D superconductivity \cite{berg07,steg13}.   

The surprising behavior occurs for A2, where $R_s$ virtually plateaus at $R_Q$ for $H\sim H_{\rm 2D}$; however, it is also a consistent response for a 2D superconductor in a strong magnetic field.  The field penetrates the sample as magnetic flux quanta that are screened by vortices of superconducting current.  If the vortices are pinned by a combination of quenched disorder and electromagnetic interactions between layers \cite{steg13,rama09}, then $R_s$ will be zero (as observed for A1); on the other hand, if the vortices are not pinned in one part of the sample, dissipation will be observed (as for A2).  In a model for the field-driven superconductor-insulator transition in disordered 2D superconductors, a boson-vortex duality has been proposed, which predicts that $R_s=R_Q$ when the vortices can flow freely, provided that $R_{\rm H}=0$ \cite{fish90b,brez16} as in the present case.

Of course, the quantized vortices are only defined when a locally-coherent supercurrent is present.  A 2D superconductor can only be ordered in the limit of zero current \cite{fish91}; with a finite current, we weaken the superconducting correlations and introduce a finite resistance.  To the extent that the boson-vortex duality applies, we may expect a complementary effect on the vortices.  We demonstrate this complementarity in Fig.~4A and B.  At $T=5$~K, $R_s$ for A1 is already finite in the 2D superconducting regime, and we observe that it increases with the measurement current.  In contrast, for A2, $R_s$ {\it decreases} as the current is raised.  This effect is also clearly demonstrated in Fig.~4C, which compares the current-dependent $R_s$ for A1 and A2 at $H\approx H_{\rm 2D}$ and $T\approx 1$~K: here, the changes for A1 and A2 are essentially equal and opposite.  A similar trend is demonstrated by a plot of $R_s$ at $H\approx H_{\rm 2D}$ vs.\ temperature for A1 and A2, as in Fig.~4D.

The most remarkable feature is the ultra-quantum metal phase at high field.  Looking at Fig.~3D and E, we see for A1 and A2 that the rising field eventually destroys the 2D superconducting correlations and causes a large rise in $R_s$, which then starts to saturate for $H>H_{\rm UQM}$.  (In the figure, the scaling based on $H_{\rm 2D}$ and $H_{\rm 3D}$ also collapses the curves at the transition to the UQM phase, so we define $H_{\rm UQM} \equiv H_{\rm 2D}+\Delta H$.)  From the scaling, it is clear that $R_s\rightarrow 2R_Q$ in the limit of $T\rightarrow 0$ when $H$ is sufficiently above $H_{\rm UQM}$.  Although the saturation limit is large, the fact that we approach saturation is consistent with a metallic state within the CuO$_2$ planes.  This state is also unusual for a metal in that $R_H\approx0$.  We note that related results have been observed in a study of Eu-doped La$_{2-x}$Sr$_x$CuO$_4$ \cite{shi18}.

Sample B shows complementary behavior, with $R_s$ decreasing substantially above $H_{\rm UQM}$.  Loss of the 2D superconducting order leads to a dramatic reduction in the magnitude of the $\rho_c$ contribution to $R_s$ for sample B.  Based on the $\rho_c$ contribution to B, we estimate that $\rho_c/\rho_{ab}$ drops to $<10^2$ at $T=3$~K (details in SM), compared to $10^4$ at zero field and $T=40$~K.   We also note for sample B that $R_s$ depends on the measurement current for $H> H_{\rm 2D}$, as shown in Fig.~S5.

\paragraph*{DISCUSSION} \ \ 

What do these results tell us about the UQM phase?  The standard theoretical description of a metal is in terms of fermionic quasiparticles.  To estimate the transport properties due to fermionic excitations at low temperature after suppression of superconductivity, we can extrapolate from the zero-field normal state response previously reported for $T>40$~K \cite{li07,tran08}.  In particular, $\rho_c$ has a large, nonmetallic magnitude that grows with cooling, and extrapolates to insulating behavior.  If we assume that quasiparticles could explain our observations for A1 and A2, we would then expect to have $\rho_c/\rho_{ab} > 10^4$, inconsistent with our results for B.  In-plane quasiparticles should also yield a finite value of $R_{\rm H}$, as observed in YBa$_2$Cu$_3$O$_{6+x}$ at high field \cite{lebo07}, which is, again, contrary to our results for LBCO. 

The Hall constant is zero in the superconducting state because of the particle-hole symmetry of the Bogoliubov quasiparticles that make up the Cooper pairs.  The observation that $R_{\rm H}$ remains negligible at high field suggests that such symmetry may survive.  Such behavior (zero Hall resistivity and finite longitudinal resistivity) has been observed previously in disordered 2D superconductors \cite{brez17}, and hence our observations are not entirely unique.  Nevertheless, the longitudinal resistivity in that case is much smaller than in the normal state, corresponding to the recently discussed  ``anomalous metal regime'' \cite{kapi19}.  Our situation is different, with an in-plane metallic resistance much larger than in the normal state at, for example, $T=50$~K, and hence it requires a distinct interpretation.  [An alternative fermionic picture, to be discussed elsewhere, involves precisely compensating hole and electron pockets of quasiparticles; however, angle-resolved photoemission studies provide no evidence of such pockets in zero field \cite{vall06,he09}.]

It was noted quite some time ago that superconducting order in underdoped cuprates may be limited by phase coherence and not by electron pairing \cite{emer95a}.  In the present case, we have destroyed the superconducting phase coherence with the magnetic field, but  electron pairs may survive.  It has also been proposed, in a model based on coupled charge stripes, that strong electron-pair correlations might live within the charge stripes, with pair hopping between stripes providing the phase coherence \cite{emer97,carl00}.  The large magnitude of $R_s$ in the UQM state is consistent with localization of surviving electron pairs within charge stripes; nevertheless, incoherent tunneling of bosonic pairs between stripes could yield finite conductivity (failed insulator).  Regarding the $c$-axis response inferred from sample B, the large magnitude of $R_s$ in the 2D regime is consistent with PDW order, which results in cancellation of interlayer pair tunneling \cite{berg07,raja18} (failed 3D superconductor); with the loss of that order, even limited tunneling of pairs between layers would reduce $\rho_c/\rho_{ab}$ in the UQM phase, as appears to happen.  The nonlinear transport in sample B (see Fig.~S5) is also suggestive of pairing correlations that can be destroyed by high current density.  

The experimental results leave us with the implication that the UQM phase is a Bose metal.   There have been various theoretical proposals of a Bose metal state, but applications to real materials have so far been confounded by coexisting fermionic quasiparticles \cite{kapi19}.  In LBCO at high field, quasiparticles do not appear to be relevant.  While further experimental and theoretical work is needed, we believe that our circumstantial evidence supports the perspective that charge stripes in cuprates, including dynamic stripes, are good for electron pairing \cite{emer97,zaan01,whit09,corb14,huan17,carr02}, even if stripe ordering may limit the degree of superconducting phase coherence.

\newpage

\paragraph*{MATERIALS AND METHODS}

\paragraph*{Materials and sample preparation}

 Several samples were oriented with Laue x-ray diffraction, cut from the large crystal with a wire saw, and polished with diamond sand paper down to 0.3-$\mu$m roughness, achieving an almost perfect rectangular shape for direct $ab$-plane charge transport measurements. No defects or cracks were visible on the surfaces of the samples. PELCO and Dupont 6838 silver paste and gold wires were used to contact samples in a Hall bar geometry, and samples were annealed in air flow at 450~$^\circ$C to minimize the contact resistance. The current contacts were made by covering the whole area of the two opposing ends to ensure uniform current flow, and the voltage contacts were made narrow to minimize the uncertainty in the absolute values of the resistance (inset of Fig.~2a). Detailed high field measurements were performed on two samples, A and B, with dimensions $3.85 \times 0.94 \times 0.53$ mm$^3$ and $3.08 \times 1.24 \times 0.58$ mm$^3$ ($a \times b \times c$), respectively. 

\paragraph*{Charge transport measurements}

A standard four probe configuration was used for charge transport measurements. Zero-field resistivity was measured using a helium cryostat, Signal Recovery 7265 lock-in amplifiers, Keithley 6220 current sources, and Keithley 2182A nanovolt meters. 

The measurements as a function of magnetic field were performed with the 35 Tesla resistive magnet and a $^3$He cryostat at the DC Field Facility, National High Magnetic Field Laboratory. A straight probe was used to measure two samples at the same time. Signal Recovery 7265 and Stanford Research 865A lock-in amplifiers, Keithley 2182A nanovolt meters, and Lake Shore 372 ac resistance bridges were used to measure magnetoresistivity and Hall coefficient. Magnetoresistivity measurements were done with low frequency ac current at 31.6~$\mu$A for 0.35 K~$\le T < 5$~K and with currents of 31.6, 100, 316 $\mu$A for 5~K~$\le T \le 50$~K. Distinct frequencies were used (13 Hz for sample A, 16 Hz for sample B) to avoid crosstalk.  Hall measurements were done with currents of 31.6~$\mu$A for $T < 25$~K and 100~$\mu$A for $T \ge 25$~K to minimize heating effects at low temperatures and the noise level at high temperatures.   All measurements were performed by fixing the temperature and sweeping the field, with the sweep rate varying from 1~T/min at low temperatures to 3~T/min at high temperatures.

ac $dV/dI$  measurements of nonlinear resistivity were performed with Signal Recovery 7265 and Stanford Research 865A lock-in amplifiers, Keithley 2182A nanovolt meters, and DL Instruments 1211 current preamplifers, with an instrumentation set-up similar to that used in a previous work \cite{shi18}.  ac voltage across the sample was measured by applying a dc current bias (10 -- 200~$\mu$A, provided by Keithly 6221 current sources) and a small ac current excitation ($\sim 10\ \mu$A) at 13 and 16~Hz.  Joule heating was negligible at high temperatures (Fig.~4) and relatively small compared to nonlinear resistivity signals at low temperatures. $dV/dI$ measurements were performed at 0.4~K~$\le T \le 5$~K.

\paragraph*{Supplementary materials}  \ \  \\
Estimation of $c$-axis misorientation for sample B.\\
Fig. S1.  Temperature-magnetic field phase diagram for A2.\\
Fig. S2.  Field-dependence of Hall voltage at various temperatures.\\
Fig. S3.  Sheet resistance at constant magnetic field.\\
Fig. S4.  ac nonlinear resistivity at $H_{\rm 2D}$ for various temperatures.\\
Fig. S5.  Additional results for sample B.\\
Fig. S6.  Comparison between A1 and previous results for a similar sample.

%BibTeX users: After compilation, comment out the following two lines and paste in
% the generated .bbl file. 

%\bibliography{scibib}

%\bibliographystyle{Science}
%\bibliography{LNO,theory,neutrons,misc}

\begin{thebibliography}{10}

\bibitem{keim15}
B.~Keimer, S.~A. Kivelson, M.~R. Norman, S.~Uchida, J.~Zaanen, ``From quantum matter to
  high-temperature superconductivity in copper oxides,'' {\it Nature\/}
  {\bf 518}, 179--186 (2015).

\bibitem{oren00}
J.~Orenstein, A.~J. Millis, ``Advances in the
  Physics of High-Temperature Superconductivity,'' {\it Science\/} {\bf 288}, 468--474 (2000).

\bibitem{frad15}
E.~Fradkin, S.~A. Kivelson, J.~M. Tranquada, ``Colloquium: Theory of intertwined orders in high  temperature superconductors,'' {\it Rev. Mod. Phys.\/} {\bf 87},
  457--482 (2015).

\bibitem{zhen17}
B.-X. Zheng, C.-M. Chung, P. Corboz, G. Ehlers, M.-P. Qin, R. M. Noack, H. Shi, S. R. White, S. Zhang, G. Kin-Lac Chan, ``Stripe order in the underdoped region of the
  two-dimensional Hubbard model,'' {\it Science\/} {\bf 358}, 1155--1160 (2017).

\bibitem{huan17}
E.~W. Huang, C. B. Mendl, S. Liu, S. Johnston, H.-C. Jiang, B. Moritz, T. P. Devereaux, ``Numerical
  evidence of fluctuating stripes in the normal state of high-$T_c$ cuprate
  superconductors,'' {\it Science\/} {\bf 358}, 1161--1164 (2017).

\bibitem{whit09}
S.~R. White, D.~J. Scalapino, ``Pairing on striped $t$-$t'$-$J$ lattices,'' {\it Phys. Rev. B\/} {\bf 79}, 220504 (2009).

\bibitem{corb14}
P.~Corboz, T.~M. Rice, M.~Troyer, ``Competing States in the $t$-$J$ Model: Uniform $d$-Wave State versus Stripe
  State,'' {\it Phys. Rev. Lett.\/} {\bf 113}, 046402
  (2014).

\bibitem{huck11}
M.~H\"ucker, M. v. Zimmermann, G. D. Gu, Z. J. Xu, J. S. Wen, G. Y. Xu, H. J. Kang, A. Zheludev, J. M. Tranquada, ``Stripe order in superconducting
  La$_{2-x}$Ba$_{x}$CuO$_{4}$ ($0.095\le{}x\le{}0.155$),''  {\it Phys. Rev. B\/} {\bf 83}, 104506 (2011).

\bibitem{carl00}
E.~W. Carlson, D.~Orgad, S.~A. Kivelson, V.~J. Emery, ``Dimensional crossover in
  quasi-one-dimensional and high-$T_c$ superconductors,'' {\it Phys. Rev. B\/} {\bf
  62}, 3422--3437 (2000).

\bibitem{seba15}
S.~E. Sebastian, C.~Proust, ``Quantum Oscillations in Hole-Doped Cuprates,'' {\it Annu. Rev. Condens. Matter Phys.\/} {\bf 6},
  411--430 (2015).

\bibitem{huck13}
M.~H\"ucker, M. v. Zimmermann, Z. J. Xu, J. S. Wen, G. D. Gu, J. M. Tranquada, ``Enhanced charge stripe order of
  superconducting La$_{2-x}$Ba$_{x}$CuO$_{4}$ in a magnetic field,'' {\it Phys. Rev. B\/} {\bf 87}, 014501 (2013).

\bibitem{kim08b}
J.~Kim, A.~Kagedan, G.~D. Gu, C.~S. Nelson, Y.-J. Kim, ``Magnetic field dependence of charge
  stripe order in La$_{2-{}x}$Ba$_{x}$CuO$_{4}$ $(x\approx{}\frac{1}{8})$,'' {\it Phys. Rev. B\/}
  {\bf 77}, 180513 (2008).

\bibitem{gerb15}
S.~Gerber, H. Jang, H. Nojiri, S. Matsuzawa, H. Yasumura, D. A. Bonn, R. Liang, W. N. Hardy, Z. Islam, A. Mehta, S. Song, M. Sikorski, D. Stefanescu, Y. Feng, S. A. Kivelson, T. P. Devereaux, Z.-X. Shen, C.-C. Kao, W.-S. Lee, D. Zhu, J.-S. Lee, ``Three-dimensional charge density wave order in YBa$_2$Cu$_3$O$_{6.67}$ at
  high magnetic fields,'' {\it Science\/} {\bf 350}, 949--952 (2015).

\bibitem{li07}
Q.~Li, M.~{H\"ucker}, G.~D. Gu, A.~M. Tsvelik, J.~M. Tranquada, ``Two-Dimensional Superconducting
  Fluctuations in Stripe-Ordered La$_{1.875}$Ba$_{0.125}$CuO$_4$,'' {\it Phys. Rev.
  Lett.\/} {\bf 99}, 067001 (2007).

\bibitem{hime02}
A.~Himeda, T.~Kato, M.~Ogata, ``Stripe States with Spatially
  Oscillating $d$-Wave Superconductivity in the Two-Dimensional $t-t'{}-J$
  Model,''  {\it Phys. Rev. Lett.\/} {\bf 88}, 117001 (2002).

\bibitem{berg07}
E.~Berg, E. Fradkin, E.-A. Kim, S. A. Kivelson, V. Oganesyan, J. M. Tranquada, S. C. Zhang, ``Dynamical Layer Decoupling in a Stripe-Ordered High-$T_c$
  Superconductor,'' {\it Phys. Rev. Lett.\/} {\bf 99}, 127003 (2007).

\bibitem{steg13}
Z.~Stegen, S. J. Han, J. Wu, A. K. Pramanik, M. H\"ucker, G. D. Gu, Q. Li, J. H. Park, G. S. Boebinger, J. M. Tranquada, ``Evolution of
  superconducting correlations within magnetic-field-decoupled
  La$_{2-x}$Ba$_{x}$CuO$_{4}$ ($x=0.095$),'' {\it Phys. Rev. B\/} {\bf 87}, 064509 (2013).

\bibitem{fuji04}
M.~Fujita, H.~Goka, K.~Yamada, J.~M. Tranquada, L.~P. Regnault, ``Stripe order, depinning, and
  fluctuations in La$_{1.875}$Ba$_{0.125}$CuO$_4$ and
  La$_{1.875}$Ba$_{0.075}$Sr$_{0.050}$CuO$_4$,'' {\it Phys. Rev.
  B\/} {\bf 70}, 104517 (2004).

\bibitem{tana98}
H.~Tanabe, S.~Watauchi, I.~Tanaka, H.~Kojima, ``TSFZ Growth of La$_{2-x}$Ba$_x$CuO$_4$ Single Crystals Under Varied Oxygen Partial Pressure,'' {\it Advances in
  Superconductivity X\/}, K.~Osamura, I.~Hirabayashi, eds. (Springer, Tokyo,
  1998), p. 371--374.

\bibitem{tran04}
J.~M. Tranquada, H. Woo, T. G. Perring, H. Goka, G. D. Gu, G. Xu, M. Fujita, K. Yamada, ``Quantum magnetic
  excitations from stripes in copper oxide superconductors,'' {\it Nature\/} {\bf 429}, 534--538 (2004).

\bibitem{tran08}
J.~M. Tranquada, G. D. Gu, M. H\"ucker, Q. Jie, H.-J. Kang, R. Klingeler, Q. Li, N. Tristan, J. S. Wen, G. Y. Xu, Z. J. Xu, J. Zhou, M. v. Zimmermann, ``Evidence for unusual superconducting
  correlations coexisting with stripe order in
  La$_{1.875}$Ba$_{0.125}$CuO$_4$,'' {\it Phys. Rev. B\/} {\bf 78}, 174529 (2008).

\bibitem{home12}
C.~C. Homes, M. H\"ucker, Q. Li, Z. J. Xu, J. S. Wen, G. D. Gu, J. M. Tranquada, ``Determination of the optical properties of La$_{2-x}$Ba$_{x}$CuO$_{4}$ for
  several dopings, including the anomalous $x=\frac{1}{8}$ phase,'' {\it Phys. Rev. B\/} {\bf 85}, 134510 (2012).

\bibitem{boll11}
A.~T. Bollinger, G. Dubuis, J. Yoon, D. Pavuna, J. Misewich, I. Bozovic, ``Superconductor-insulator transition in La$_{2-x}$Sr$_x$CuO$_4$ at
  the pair quantum resistance,'' {\it Nature\/} {\bf 472}, 458--460 (2011).

\bibitem{rama09}
K.~S. Raman, V.~Oganesyan, S.~L. Sondhi, ``Biot-Savart correlations in
  layered superconductors,'' {\it Phys. Rev. B\/} {\bf 79}, 174528
  (2009).

\bibitem{fish90b}
M.~P.~A. Fisher, ``Quantum phase transitions in disordered two-dimensional
  superconductors,'' {\it Phys. Rev. Lett.\/} {\bf 65}, 923--926 (1990).

\bibitem{brez16}
N.~P. Breznay, M.~A. Steiner, S.~A. Kivelson, A.~Kapitulnik, ``Self-duality and a Hall-insulator phase near the
  superconductor-to-insulator transition in indium-oxide films,'' {\it Proc. Natl.
  Acad. Sci. USA\/} {\bf 113}, 280--285 (2016).

\bibitem{fish91}
D.~S. Fisher, M.~P.~A. Fisher, D.~A. Huse, ``Thermal fluctuations, quenched disorder, phase transitions, and transport
  in type-II superconductors,'' {\it Phys. Rev. B\/} {\bf 43}, 130--159
  (1991).

\bibitem{shi18}
Z.~{Shi}, P.~G. {Baity}, T.~{Sasagawa}, D.~{Popovi{\'c}}, ``Unveiling the phase
  diagram of a striped cuprate at high magnetic fields: Hidden order of Cooper
  pairs,'' https://arxiv.org/abs/1801.06903 (2018).

\bibitem{lebo07}
D.~LeBoeuf, N. Doiron-Leyraud, J. Levallois, R. Daou, J.-B. Bonnemaison, N. E. Hussey, L. Balicas, B. J. Ramshaw, R. Liang, D. A. Bonn, W. N. Hardy, S. Adachi, C. Proust, L. Taillefer, ``Electron pockets in the
  fermi surface of hole-doped high-$T_c$ superconductors,'' {\it Nature\/} {\bf 450}, 533--536 (2007).

\bibitem{brez17}
N.~P. Breznay, A.~Kapitulnik, ``Particle-hole symmetry reveals failed superconductivity in the metallic
  phase of two-dimensional superconducting films,'' {\it Sci. Adv.\/} {\bf 3} (2017) 10.1126/sciadv.1700612.

\bibitem{kapi19}
A.~Kapitulnik, S.~A. Kivelson, B.~Spivak, ``Colloquium:
  Anomalous metals: Failed superconductors,'' {\it Rev. Mod. Phys.\/} {\bf 91},
  011002 (2019).

\bibitem{vall06}
T.~Valla, A.~V. Federov, J.~Lee, J.~C. Davis, G.~D. Gu, ``The Ground State of the Pseudogap in Cuprate
  Superconductors,'' {\it Science\/} {\bf
  314}, 1914--1916 (2006).

\bibitem{he09}
R.-H. He, K. Tanaka, S.-K. Mo, T. Sasagawa, M. Fujita, T. Adachi, N. Mannella, K. Yamada, Y. Koike, Z. Hussain, Z.-X. Shen, ``Energy gaps in the failed high-$T_c$
  superconductor La$_{1.875}$Ba$_{0.125}$CuO$_4$,'' {\it Nat. Phys.\/} {\bf 5}, 119--123 (2009).

\bibitem{emer95a}
V.~J. Emery, S.~A. Kivelson, ``Importance of
  phase fluctuations in superconductors with small superfluid density,'' {\it Nature\/} {\bf 374}, 434--437 (1995).

\bibitem{emer97}
V.~J. Emery, S.~A. Kivelson, O.~Zachar, ``Spin-gap proximity effect
  mechanism of high-temperature superconductivity,'' {\it Phys. Rev. B\/} {\bf 56}, 6120--6147
  (1997).

\bibitem{raja18}
S.~Rajasekaran, J. Okamoto, L. Mathey, M. Fechner, V. Thampy, G. D. Gu, A. Cavalleri, ``Probing optically silent superfluid stripes in
  cuprates,'' {\it Science\/} {\bf 359}, 575--579 (2018).

\bibitem{zaan01}
J.~Zaanen, O.~Y. Osman, H.~V. Kruis, Z.~Nussinov, J.~Tworzyd{\l}o, ``The geometric order of stripes and
  luttinger liquids,'' {\it Phil.
  Mag. B\/} {\bf 81}, 1485--1531 (2001).

\bibitem{carr02}
S.~T. Carr, A.~M. Tsvelik, ``Superconductivity and charge-density waves in a quasi-one-dimensional
  spin-gap system,'' {\it Phys. Rev. B\/} {\bf 65}, 195121 (2002).

\end{thebibliography}

%\end{document}

\paragraph*{Acknowledgments:}
The authors are grateful to E. S. Choi and S. Maier for assistance with the experiment, and to E. Berg, M. S. Foster, E. Fradkin, S. A. Kivelson, and A. Sapkota for valuable discussions and comments.  {\bf Funding:} Work at Brookhaven is supported by the Office of Basic Energy Sciences, Materials Sciences and Engineering Division, U.S. Department of Energy (DOE) under Contract No.\ DE-SC0012704.  The work at the National High Magnetic Field Laboratory and Florida State University was supported by NSF Grant No.\ DMR-1707785.  A portion of this work was performed at the National High Magnetic Field Laboratory, which is supported by National Science Foundation Cooperative Agreement No. DMR-1644779 and the State of Florida.
{\bf Author contributions:} J.M.T., Q.L., and D.P. conceived the experiment.  G.D.G. grew the crystals.  Y.L., J.T., and P.G.B. performed the experiments.  Y.L., D.P, A.M.T., and J.M.T. wrote the paper, with input from all authors.
{\bf Competing interests:} The authors declare that they have no competing interests.
{\bf Data and materials availability:} All data needed to evaluate the conclusions in the paper are present in the paper and/or the Supplementary Materials.  Additional data available from authors upon request.

%Here you should list the contents of your Supplementary Materials -- below is an example. 
%You should include a list of Supplementary figures, Tables, and any references that appear only in the SM. 
%Note that the reference numbering continues from the main text to the SM.
% In the example below, Refs. 4-10 were cited only in the SM.     

% For your review copy (i.e., the file you initially send in for
% evaluation), you can use the {figure} environment and the
% \includegraphics command to stream your figures into the text, placing
% all figures at the end.  For the final, revised manuscript for
% acceptance and production, however, PostScript or other graphics
% should not be streamed into your compliled file.  Instead, set
% captions as simple paragraphs (with a \noindent tag), setting them
% off from the rest of the text with a \clearpage as shown  below, and
% submit figures as separate files according to the Art Department's
% instructions.

\clearpage

\begin{figure}[h]
 \centering
   \includegraphics[width=0.5\columnwidth]{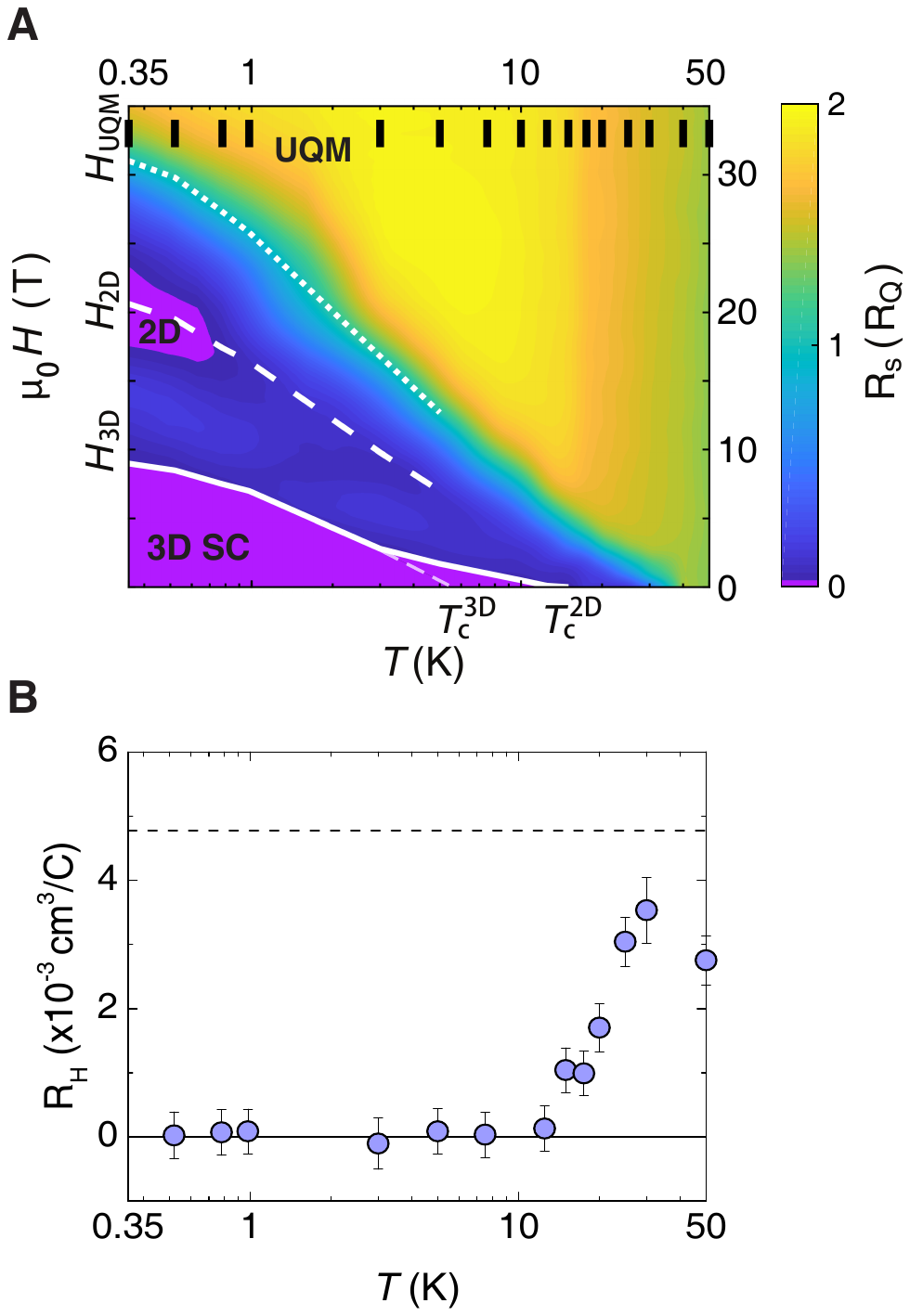}
    \caption{\label{fg:phase} {\bf Phase diagram of La$_{2-x}$Ba$_x$CuO$_4$ with $x=0.125$ in terms of sheet resistance and Hall coefficient.} (A) Interpolated color contour plot of the sheet resistance $R_s$ as a function of temperature and magnetic field. Black vertical marks indicate measurement temperatures. The regimes of 3D and 2D superconductivity with zero electrical resistance are labeled; the ultra-quantum metal phase occurs at fields above the dotted line. Characteristic fields $H_{\rm 3D}$, $H_{\rm 2D}$, and $H_{\rm UQM}$ (defined in the text) are over-plotted as solid, dashed, and dotted white lines, respectively.  (These results are for sample A1; for an analogous plot for sample A2, see Fig. S1.) (B) Hall coefficient as a function of temperature, with error bars obtained by averaging over the entire field range (0 to 35~T, see Fig.~S2) at each temperature.  $R_{\rm H}$ is effectively zero below 15~K, as expected for a superconductor, and it rises to the normal-state magnitude around $\sim40$~K. The upper dashed line indicates the magnitude of $R_H$ that would be expected in a one-band system with a nominal hole density of 0.125.  Results from sample A, as described in the text.}
\end{figure}

\begin{figure}[h]
 \centering
    \includegraphics[width=0.5\columnwidth]{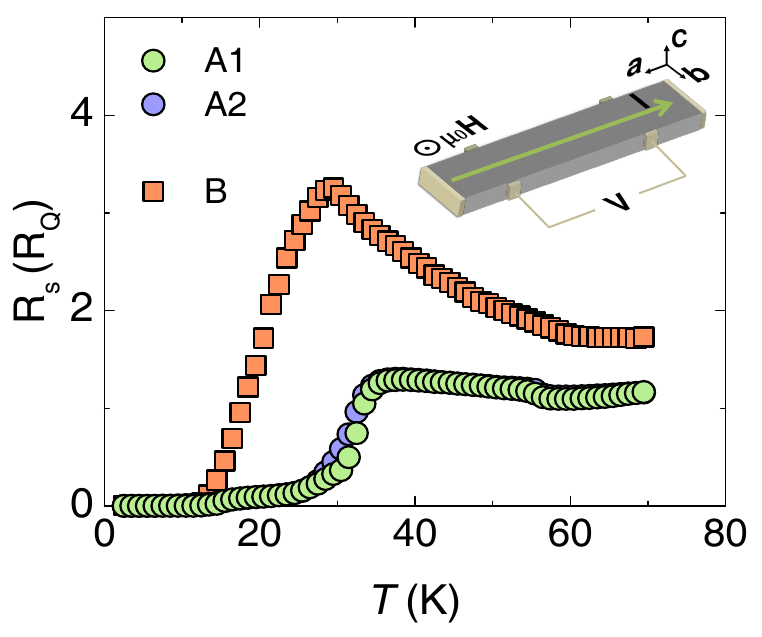}
    \caption{\label{fg:zero} {\bf Sheet resistance in zero magnetic field.} $R_s$  as a function of temperature for samples A (circles) and B (squares).  Inset shows sample and standard four-probe measurement configuration.  Sample A has voltage contacts on two edges, resulting in measurements labelled A1 and A2. The difference between samples A and B is consistent with a small $c$-axis contribution to the resistance of B due to a slight misorientation of the $c$ axis.}
\end{figure}

\begin{figure}[h]
 \centering
    \includegraphics[width=0.6\textwidth]{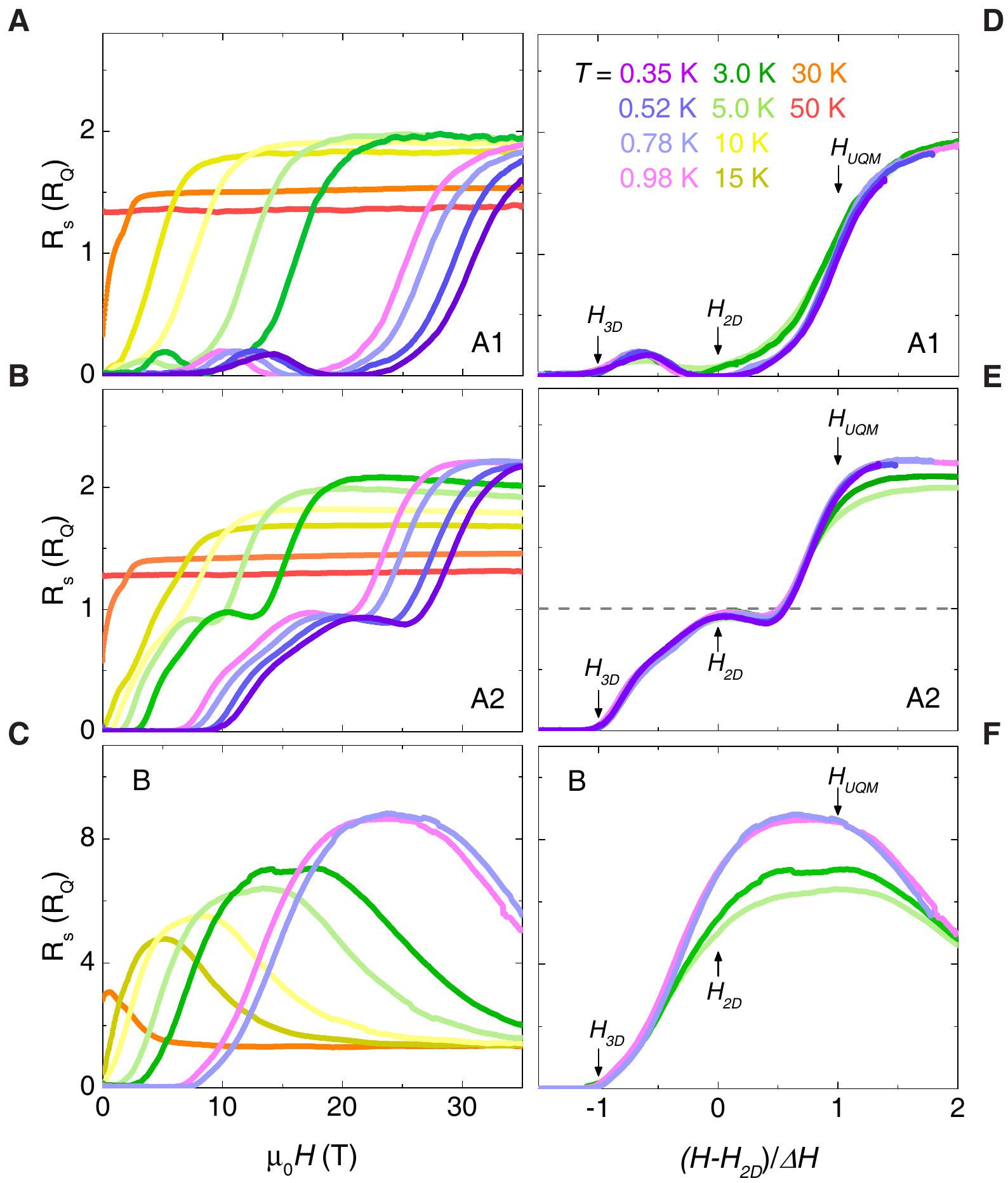}
    \caption{\label{fg:scale} {\bf Sheet resistance as a function of magnetic field.}  Results at various temperatures for samples (A) A1, (B) A2, and (C) B.  (D)-(F) show the same data plotted vs a scaled magnetic field, as discussed in the text. $H_{\rm 2D}$ corresponds to the field at which zero resistance is seen in A1 and a local maximum of $R_s \approx R_Q$ occurs in A2.  Measurement temperatures are color coded, as indicated in (D). $R_s$ was measured with an ac current of 31.6~$\mu$A.  (For plots of $R_s$ vs.\ $T$, see Fig.~S3.)}
\end{figure}

\begin{figure}[h]
 \centering
    \includegraphics[width=0.7\textwidth]{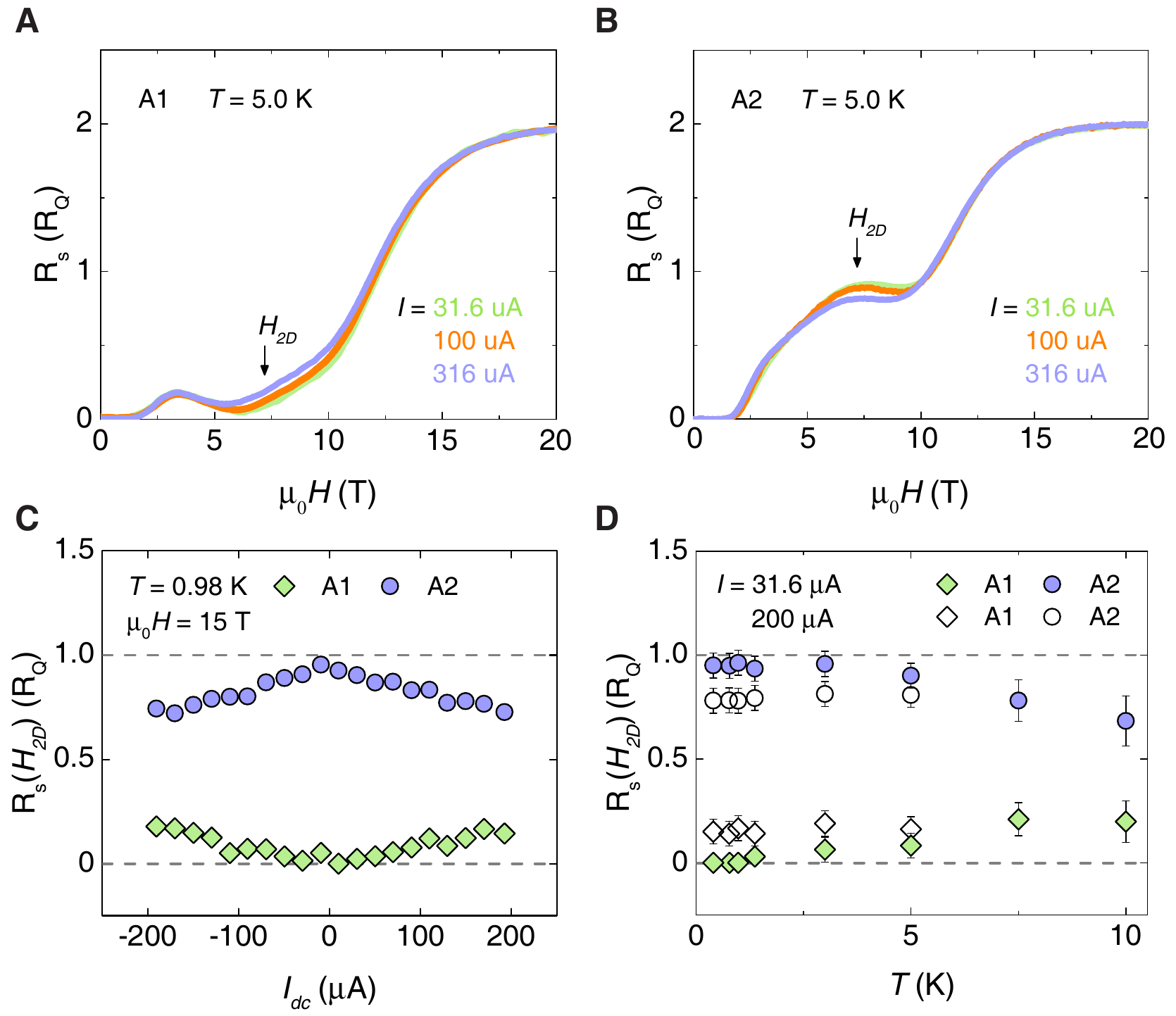}
    \caption{\label{fg:nonlin} {\bf Current dependence of sheet resistance.} Results for samples (A) A1 and (B) A2 at $T=5.0$~K with various currents. A significant difference between measurements is observed only in the 2D SC regime. (C) Current dependence of sheet resistance at $\mu_0H= 15\ \mbox{\rm T}\approx \mu_0H_{\rm 2D}$ at $T=0.98$~K for samples A1 (diamonds) and A2 (circles), obtained by nonlinear dV/dI measurements with an ac current of $\sim 10$~$\mu$A, in addition to the larger dc current (see Materials and Methods). (D)  Variation of sheet resistance at $H\approx H_{\rm 2D}$ as a function of temperature measured with dc currents of 31.6 (filled symbols) and 200 $\mu$A (open symbols).  (Further data are presented in Fig.~S4.)}
\end{figure}

\end{document}